\documentclass[12pt]{article}

\usepackage{graphicx}

\title{Characteristic time scales of tick quotes on foreign currency
 markets: an empirical study and agent-based model}
\author{Aki-Hiro Sato \\
Department of Applied Mathematics and Physics, \\
Graduate School of Informatics, Kyoto University, \\
Kyoto 606-8501, Japan}

\begin{document}
\maketitle

\begin{abstract}
Power spectrum densities for the number of tick quotes per minute
(market activity) on three currency markets (USD/JPY, EUR/USD, and
JPY/EUR) for periods from January 1999 to December 2000 are
analyzed. We find some peaks on the power spectrum densities at a few 
minutes. We develop the double-threshold agent model and confirm
that stochastic resonance occurs for the market activity of this model. 
We propose a hypothesis that the periodicities found on the power spectrum
densities can be observed due to stochastic resonance.
\end{abstract}

\noindent
tick quotes, foreign currency market, power spectrum density,
double-threshold agent model, stochastic resonance 

\section{Introduction}
\label{sec:introduction}
In the past few years there has been increasing interest in the
investigation of financial markets as complex systems in statistical
mechanics. The empirical analysis of the high frequency financial data
reveals nonstationary statistics of market fluctuations and several
mathematical models of markets based on the concept of the
nonequilibrium phenomena have been proposed~\cite{Mantegna:00,Dacrogna:00}.

Recently Mizuno {\it et al.} investigate high frequency data of
the USD/JPY exchange market and conclude that dealers' perception and
decision are mainly based on the latest 2 minutes data~\cite{Mizuno}. This
result means that there are feedback loops of information in the foreign
currency market.

As microscopic models of financial markets some agent models are
proposed~\cite{Aki:98,Lux:99,Challet:01,Jefferies:01}. Specifically 
Ising-like models are familiar to statistical physicists and have been
examined in the context of econophysics. The analogy to the 
paramagnetic-ferromagnetic transition is used to explain crashes 
and bubbles. Krawiecki {\it et al.} consider the effect
of a weak external force acting on the agents in the Ising model of the
financial market and conclude that apparently weak stimuli from outside
can have potentially effect on financial market due to stochastic
resonance~\cite{Krawiecki}. This conclusion indicates that it is
possible to observe the effect of the external stimuli from the market
fluctuations.

Motivated by their conclusion we investigate high-frequency financial
data and find a potential evidence that stochastic resonance occurs in 
financial markets. In this article the results of data analysis are
reported and the agent-based model is proposed in order to explain this
phenomenon.

\section{Analysis}
\label{sec:analysis}
We analyze tick quotes on three foreign currency markets
(USD/JPY, EUR/USD, and JPY/EUR) for periods from January 
1999 to December 2000~\cite{CQG}. This database contains time stamps,
prices, and identifiers of either ask or bid. Since generally market
participants (dealers) must indicate both ask and bid prices in foreign
currency markets the nearly same number of ask and bid offering are 
recorded in the database. Here we focus on the ask offering and 
regard the number of ask quotes per unit time (one minute) $A(t)$ as the 
market activity. The reason why we define the number of ask quotes as 
the market activity is because this quantity represents amount of 
dealers' responses to the market.

\begin{figure}[phbt]
\includegraphics[scale=0.35]{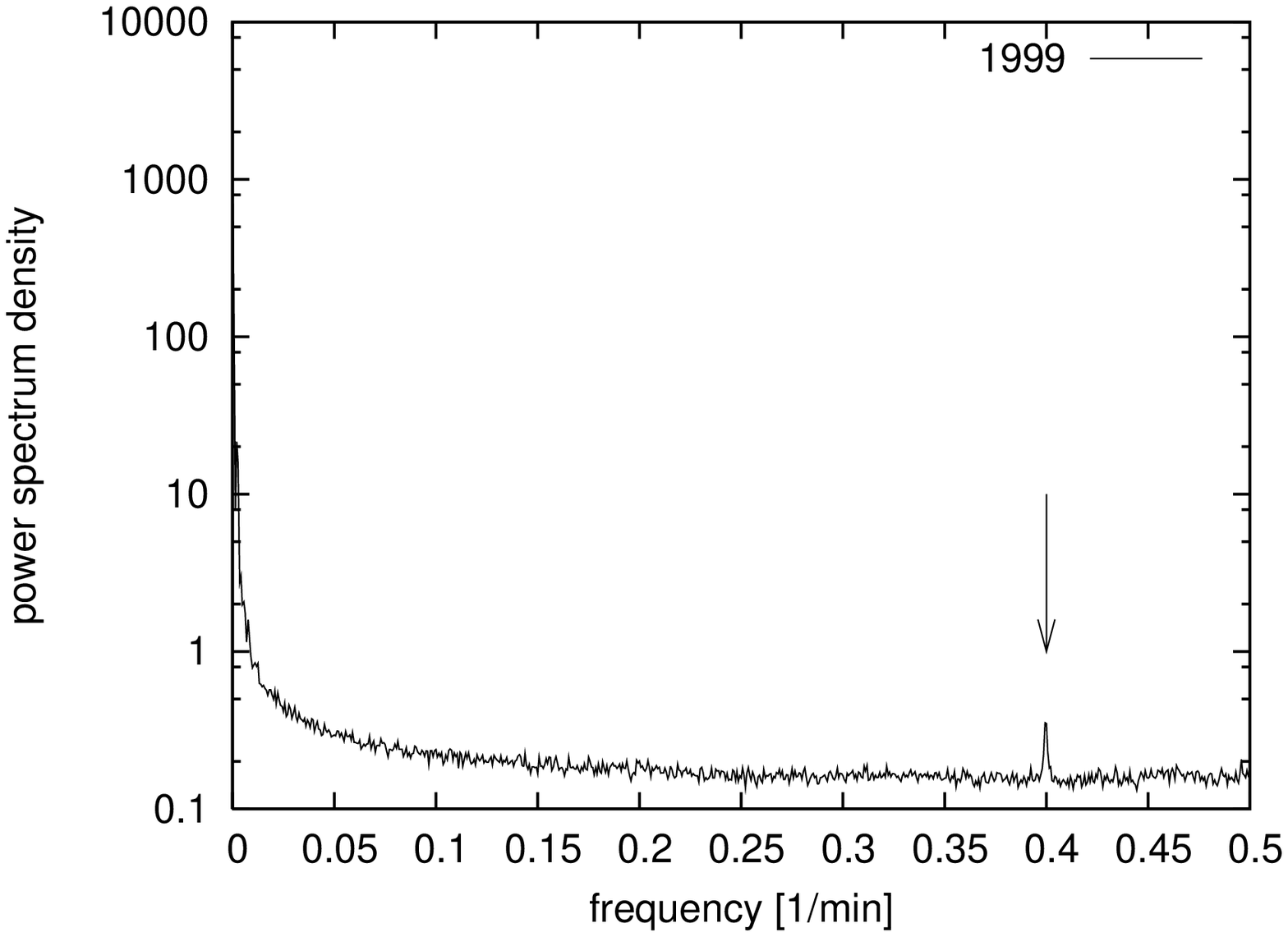}
\includegraphics[scale=0.35]{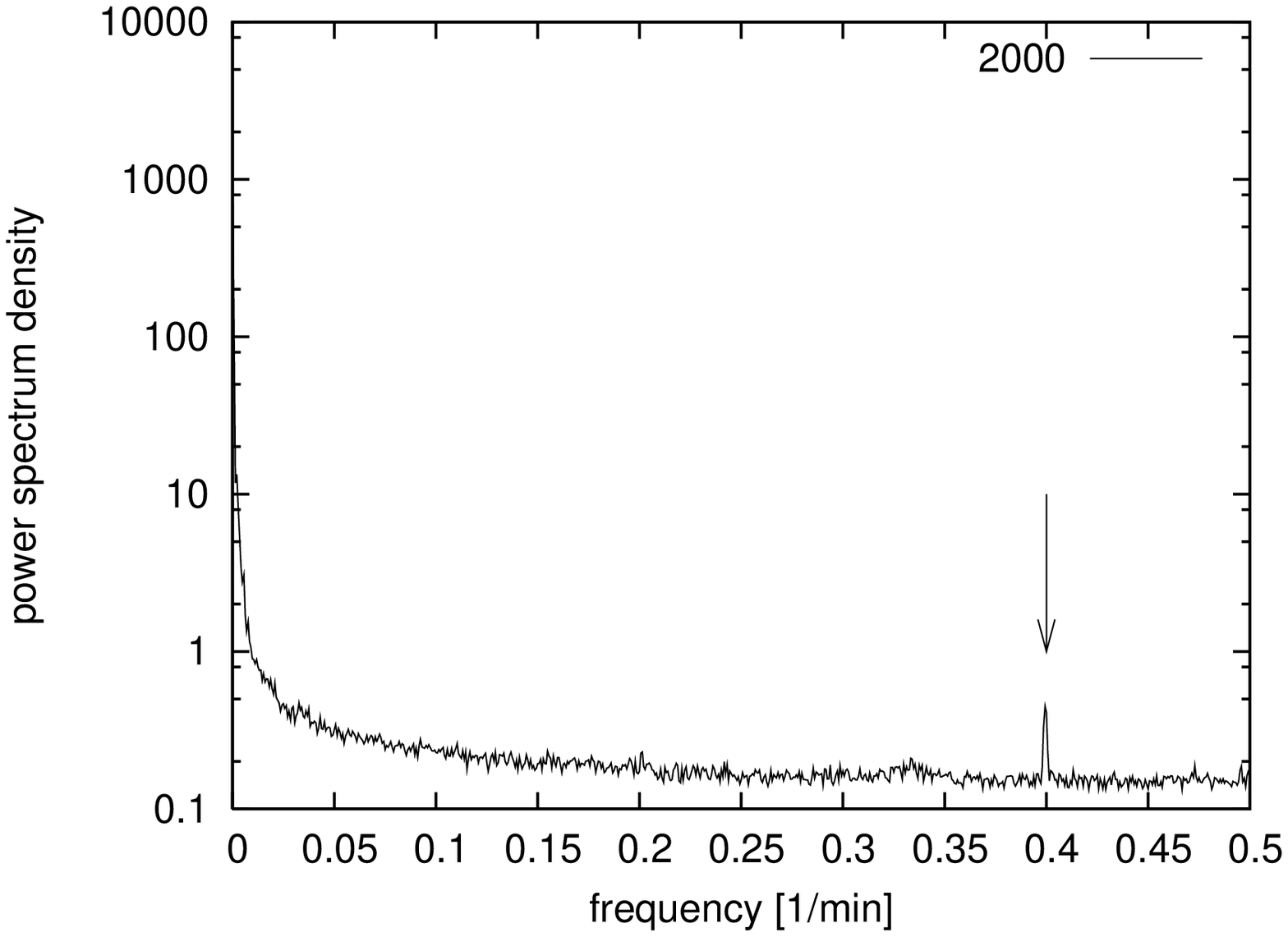}
\caption{Semi-log plots of power spectrum densities for time series of 
the number of ask quotes per minute on the USD/JPY market on 1999
 (left) and 2000 (right). These power spectrum densities are 
 estimated by averaging power spectrum densities for intraday time
 series of the number of ask quotes per minute over day for each year.} 
\label{fig:power-spectrum-usdjpy}
\end{figure}

\begin{figure}[hbt]
\includegraphics[scale=0.35]{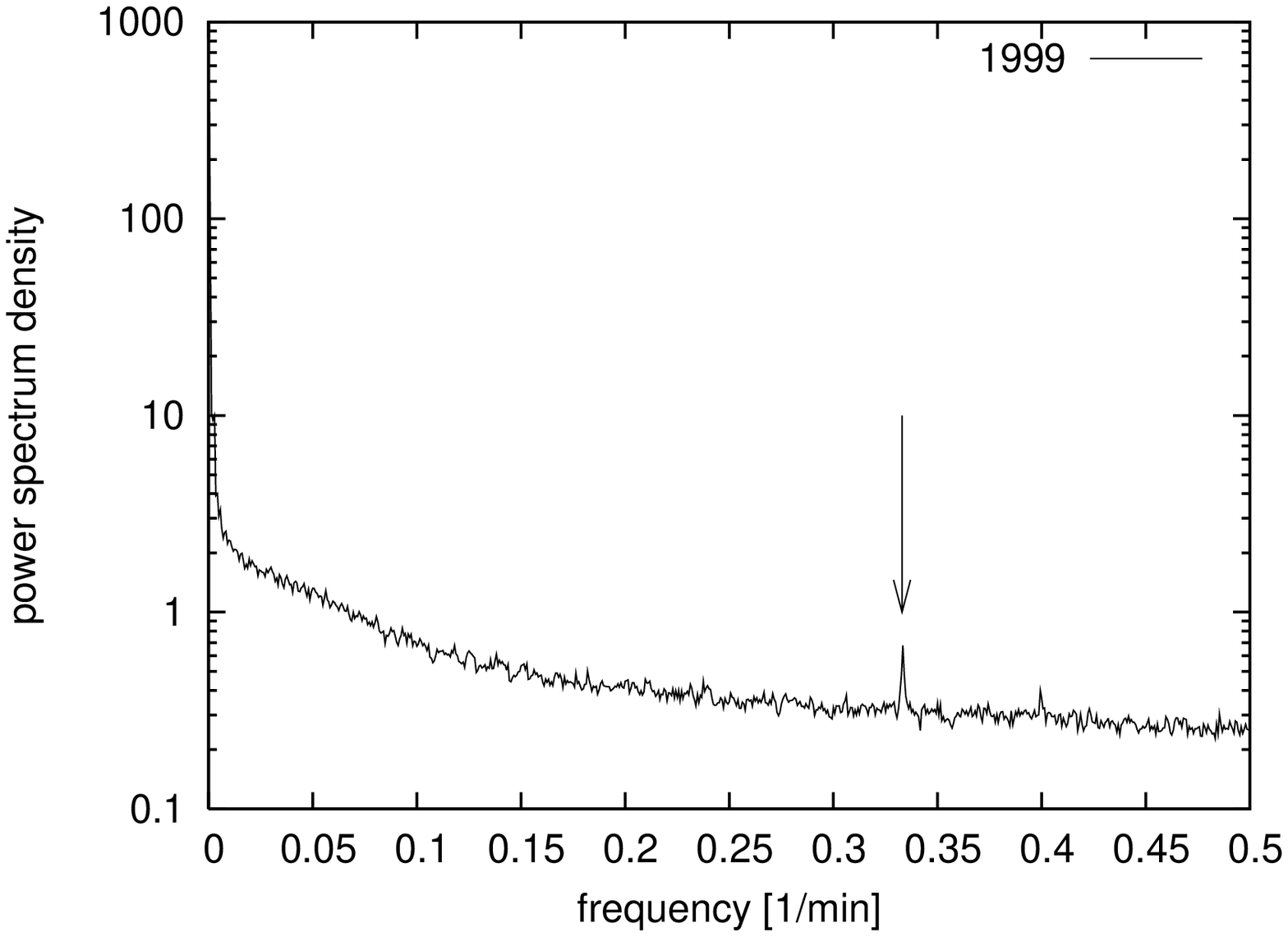}
\includegraphics[scale=0.35]{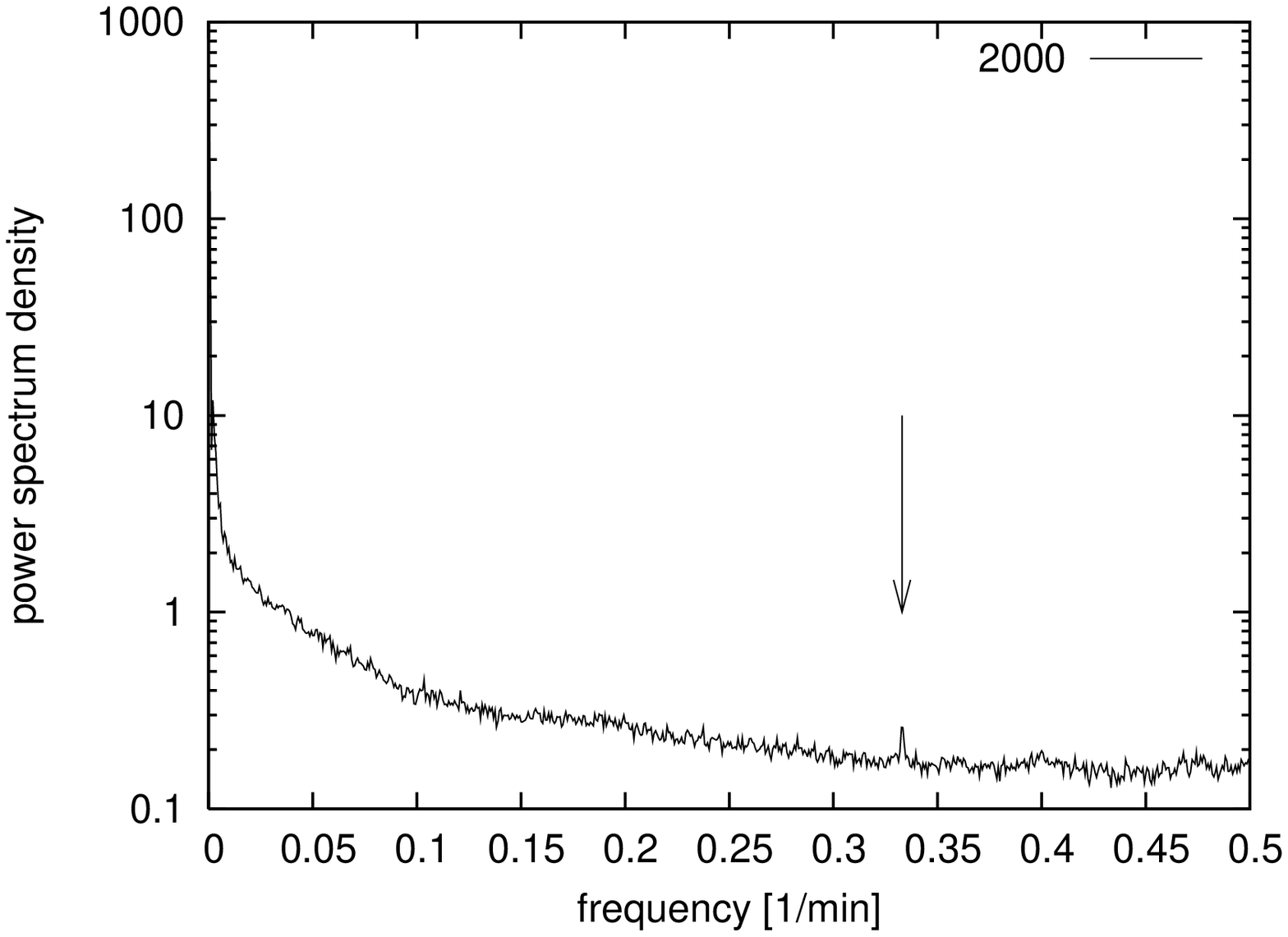}
\caption{Semi-log plots of power spectrum densities for time series of 
the number of ask quotes per minute on the EUR/USD market on 1999
 (left) and 2000 (right).}
\label{fig:power-spectrum-eurusd}
\end{figure}

\begin{figure}[hbt]
\includegraphics[scale=0.35]{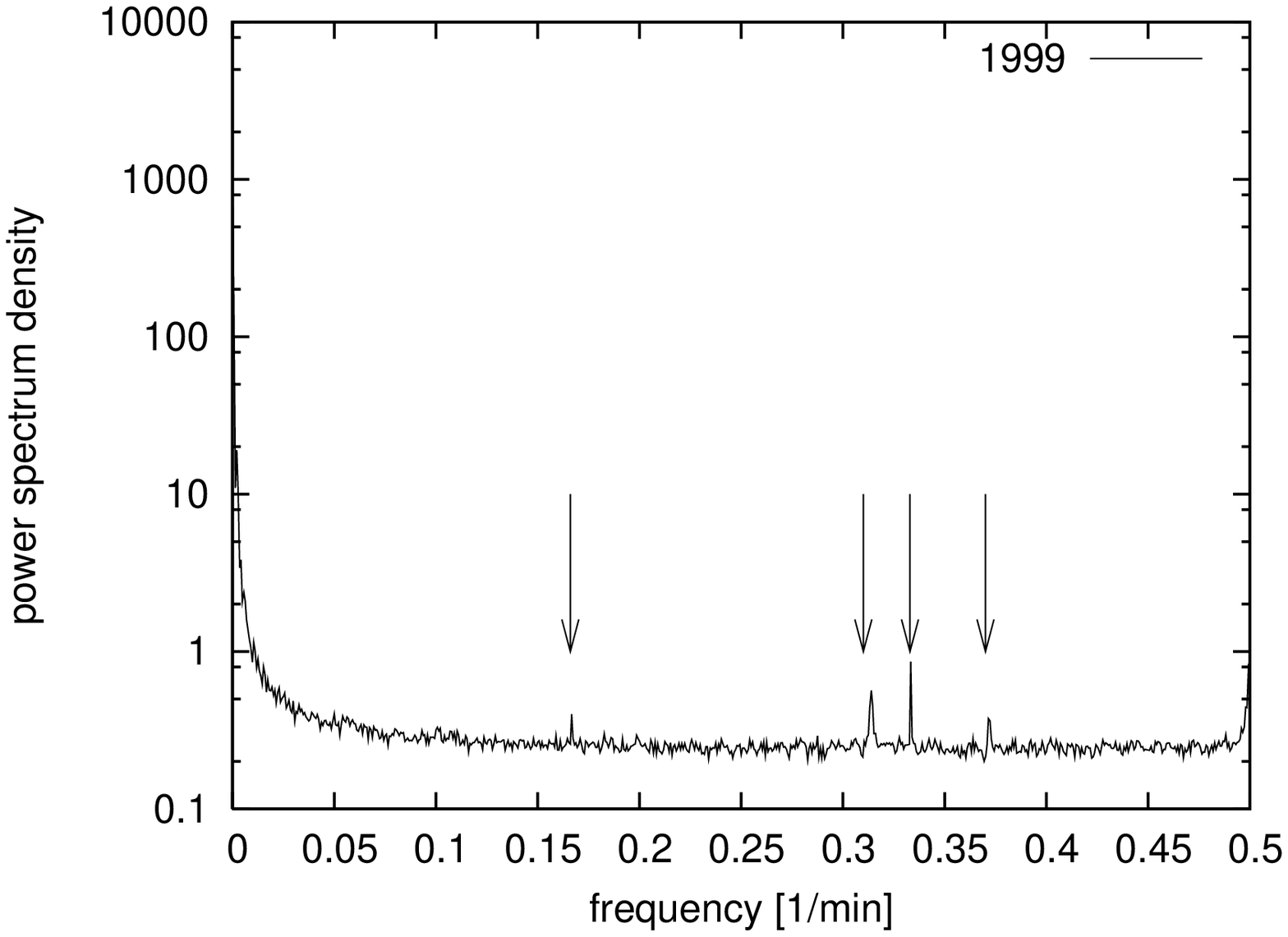}
\includegraphics[scale=0.35]{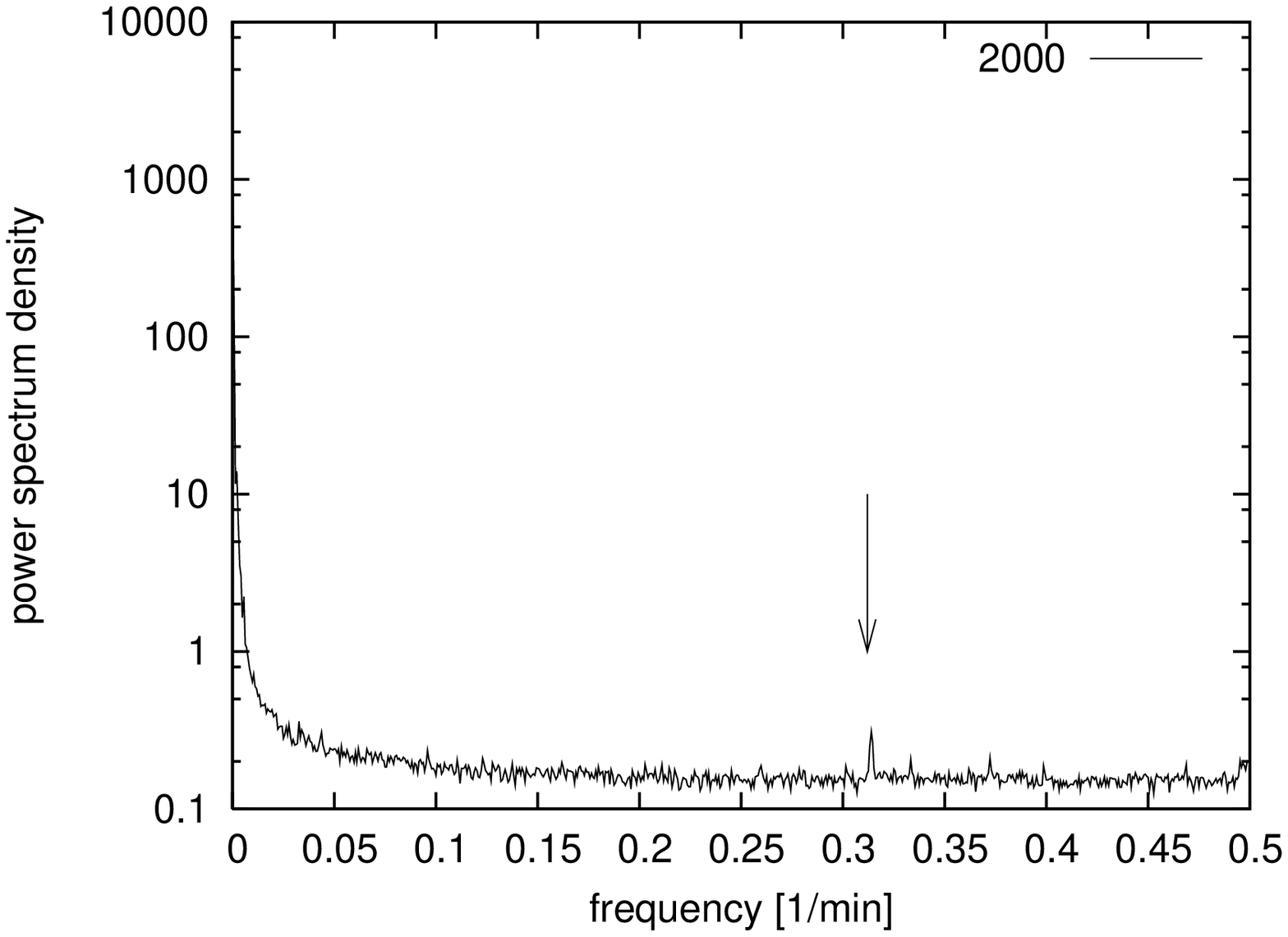}
\caption{Semi-log plots of power spectrum densities for time series of 
the number of ask quotes per minute on the EUR/JPY market on 1999
 (left) and 2000 (right).}
\label{fig:power-spectrum-eurjpy}
\end{figure}

In order to elucidate temporal structure of the market
activity power spectrum densities of $A(t)$, estimated by
\begin{equation}
S(f) =
 \frac{1}{2\pi}\lim_{T\rightarrow\infty}\frac{1}{T}
\Bigl\langle|\int_{0}^{T}A(\tau)e^{-2\pi i f \tau}d\tau|^2 \Bigr\rangle,
\end{equation}
where $f$ represents frequency, and $T$ a maximum period of the power
spectrum density, are calculated. Figs. \ref{fig:power-spectrum-usdjpy},
\ref{fig:power-spectrum-eurusd}, and \ref{fig:power-spectrum-eurjpy}
show the power spectrum densities for three foreign currency markets
(USD/JPY, UER/USD, and EUR/JPY) from January 1999 to December 2000. It
is found that they have some peaks at the high frequency region.

There is a peak at 2.5 minutes on the USD/JPY market, at 
3 minutes on the EUR/USD market, and there are some peaks on the
JPY/EUR. We confirm that these peaks appear and
disappear depending on observation periods. On the USD/JPY 
market there is the peak for periods of January 1999--July 1999,
March 2000--April 2000, and August 2000--November 2000; on the EUR/USD
market July 1999--September 1999; and on the EUR/JPY market January 
1999--March 1999, April 1999--June 1999, November 1999, and July 
2000--December 2000.

These peaks mean that market participants offer quotes periodically and
in synchronization. The possible reasons for these peaks to appear in 
the power spectrum densities of the market activity are follows: 
\begin{enumerate}
\item The market participants are affected by common periodic information. 
\item The market participants are spontaneously synchronized. 
\end{enumerate}
In the next section the double-threshold agent model is introduced and
explain this phenomenon on the basis of the reason (1). 

\section{Double-threshold agent model}
\label{sec:model}
Here we consider a microscopic model for financial markets in order to 
explain the dependency of the peak height on observation periods. We
develop the double-threshold agent model based on the threshold
dynamics. 

In foreign exchange markets the market participants attend
the markets with utilizing electrical communication devices, for example,
telephones, telegrams, and computer networks. They are driven by both 
exogenous and endogenous information and determine their investment
attitudes. Since the information to motivate buying and one of selling 
are opposite to each other we assume that the information is a scaler
variable. Moreover the market participants perceive the information
and determine their investment attitude based on the information. 
The simplest model of the market participant is an element with
threshold dynamics. 

We consider a financial market consisting of $N$ market participants
having three kinds of investment attitudes:  buying, selling, and doing
nothing. Recently we developed an array of double-threshold noisy
devices with a global feedback~\cite{Sato:05}. Applying this model to
the financial market we construct three decision model with double
thresholds and investigate the dependency of market behavior on an
exogenous stimuli.

The investment attitude of the $i$th dealer $y_i(t)$ at
time $t$ is determined by his/her recognition for the circumstances
$x_i(t) = s(t) + z_i(t)$, where $s(t)$ represents the
investment environment, and $z_i(t)$ the $i$th dealer's prediction 
from historical market data. $y_i(t)$ is given by
\begin{equation}
y_i(t) =
\left\{
\begin{array}{lll}
1  & (x_i(t)+\xi_i(t) \geq B_i(t)) & : \mbox{\it buy}\\
0  & (B_i(t) > x_i(t)+\xi_i(t) > S_i(t)) & : \mbox{\it inactive}\\
-1 & (x_i(t)+\xi_i(t) \leq S_i(t)) & : \mbox{\it sell}
\end{array}
\right.,
\end{equation}
where $B_i(t)$ and $S_i(t)$ represent threshold values to determine buying
attitude and selling attitude at time $t$, respectively. $\xi_i(t)$ is
the uncertainty of the $i$th dealer's decision-making. For simplicity it
is assumed to be sampled from an identical and independent Gaussian 
distribution,
\begin{equation}
p(\xi_i) =
\frac{1}{\sqrt{2\pi}\sigma_{\xi}}\exp\Bigl(-\frac{\xi_i^2}{2\sigma_{\xi}^2}\Bigr),  
\end{equation}
where $\sigma_{\xi} (>0)$ is a standard deviation of $\xi_i(t)$. Of
course this assumption can be weakened. Namely we can extend the uncertainty
in the case of non-Gaussian noises and even correlated noises.

The excess demand is given by the sum of investment attitudes over
the market participants, 
\begin{equation}
r(t) = N^{-1}\sum_{i=1}^{N}y_i(t),
\end{equation}
which can be an order parameter. Furthermore the market price $P(t)$ moves to the direction to the excess
demand  
\begin{equation}
\ln P(t+\Delta t) = \ln P(t) + \gamma r(t),
\end{equation}
where $\gamma$ represents a liquidity constant and $\Delta t$ is a
sampling period. $r(t)$ may be regarded as an order parameter.

The dealers determine their investment attitude based on
exogenous factors (fundamentals) and endogenous factors (market price
changes). Generally speaking, the prediction of the $i$th 
dealer $z_i(t)$ is determined by a complicated strategy described as a
function with respect to historical market prices,
$F_i(s,P(t),P(t-\Delta t),\ldots)$. Following the Takayasu's first order 
approximation~\cite{Takayasu:99} we assume that $z_i(t)$ is given by
\begin{equation}
z_i(t) = a_i(t)(\ln P(t) - \ln P(t-\Delta t)) = \gamma a_i(t) r(t-\Delta
 t),
\end{equation}
where $a_i(t)$ is the $i$th dealer's response to the market price changes.

It is assumed that the dealers' response can be separated by common
and individual factors,
\begin{equation}
a_i(t) = \zeta(t) + \eta_i(t),
\end{equation}
where $\zeta(t)$ denotes the common factor, and $\eta_i(t)$ the
individual factor. Generally these factors are time-dependent and 
seem to be complicated functions of both exogenous and endogenous
variables. 

For simplicity it is assumed that these factors vary rapidly in the limit
manner. Then this model becomes well-defined in the  stochastic manner.
We assume that $\zeta(t)$ and $\eta_i(t)$ are sampled from the following
identical and independent Gaussian distributions, respectively:
\begin{eqnarray}
P_{\zeta}(\zeta) &=&
\frac{1}{\sqrt{2\pi}\sigma_{\zeta}}\exp\Bigl(-\frac{(\zeta-a)^2}{2\sigma_{\zeta}^2}\Bigr),
  \\
P_{\eta}(\eta_i) &=& \frac{1}{\sqrt{2\pi}\sigma_{\eta}}\exp\Bigl(-\frac{\eta_i^2}{2\sigma_{\eta}^2}\Bigr),
\end{eqnarray}
where $a$ represents a mean of $\zeta(t)$, $\sigma_{\zeta} (> 0)$ a standard
deviation of $\zeta(t)$, and $\sigma_{\eta}$ a standard deviation of
$\eta (> 0)$.

Since we regard the market activity as the number of tick quotes
per unit time it should be defined as the sum of dealers' actions:
\begin{equation}
q(t) = \frac{1}{N}\sum_{i=1}^N |y_i(t)|.
\end{equation}
The market activity $q(t)$ may be regarded as an order parameter.

\section{Numerical Simulation}
\label{sec:simulation}
This agent model has nine model parameters. We fix $N=100$, $B_i=0.01$,
$S_i=-0.01$, $\gamma=0.1$, $\sigma_{\eta}=0.01$, and $a=0.0$ throughout
all numerical simulations. It is assumed that an exogenous periodic
information to the market is subject to $s(t)=q_0 \sin(2\pi \Delta t f
t)$ at $q_0=0.001$, $f=0.8$ and $\Delta t = 1$. 

We calculate the signal-to-noise ratio (SNR) of the market activity as a
function of $\sigma_{\xi}$. The SNR is defined as
\begin{equation}
SNR = \log_{10}\frac{S}{N},
\end{equation}
where $S$ represents a peak height of the power spectrum density, and
$N$ noise level. 

\begin{figure}[hbt]
\begin{center}
\includegraphics[scale=0.45]{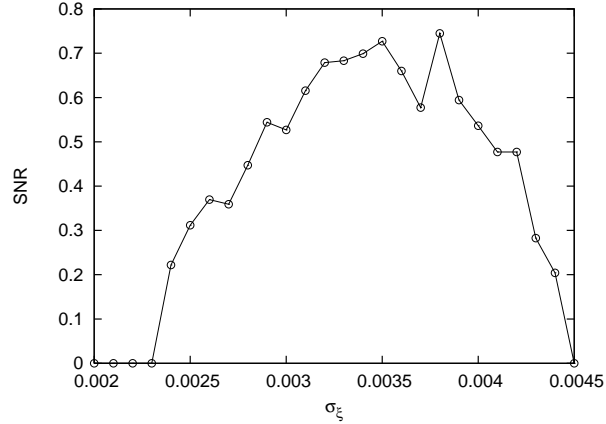}
\end{center}
\caption{Signal-to-noise ratio (SNR) obtained from power spectrum
 densities of $q(t)$ for the double-threshold agent model is plotted
 against the uncertainty of decision-making of agents at 
 $N=100$, $B_i=0.01$, $S_i=-0.01$, $\gamma=0.1$, $\sigma_{\eta} = 0.01$,
 $a=0.0$, $\sigma_{\zeta}=0.3$, $q_0=0.001$, $f=0.8$, and $\Delta t = 1$.}
\label{fig:SNR}
\end{figure}

From the numerical simulation we find non-monotonic dependency of the SNR of
$q(t)$ on $\sigma_{\xi}$. Fig. \ref{fig:SNR} shows a relation between
the SNR and the noise strength $\sigma_{\xi}$. It has
an extremal value around $\sigma_{\xi}=0.0035$. Namely the uncertainty of
decision-making plays a constructive role to enhance information 
transmission. If there are exogenous periodic information and 
the uncertainty of decision-making we can find the peak on power 
spectrum densities at appropriate uncertainty of decision-making due 
to stochastic resonance~\cite{Gammaitoni}. 

\section{Conclusion}
\label{sec:conclusion}
We analyzed time series of the number of tick quotes (market activity)
and found there are short-time periodicities in the time series. The
existence and positions of these peaks of the power spectrum densities
depend on foreign currency markets and observation periods. The power
spectrum densities have a peak at 2.5 minutes on the USD/JPY
market, 3 minutes on the EUR/USD. There are some peaks at a few minutes on
the JPY/EUR.

We developed the double-threshold agent model for financial markets where
the agents choose three kinds of states and have feedback strategies to
determine their decision affected by last price changes. From the numerical
simulation we confirmed that the information transmission is enhanced due to
stochastic resonance related to the uncertainty of decision-making of
the market participants. We propose a hypothesis that the periodicities of 
the market activity can be observed due to stochastic resonance.

Appearance and disappearance of these peaks may be related to the
efficiency of the markets. The efficiency market
hypothesis~\cite{Fama:91} says that prices reflect information. Because 
quotes make prices tick frequency can reflect information. If the
peaks of the power spectrum densities come from exogenous information 
then SNR is related to the efficiency of the market. Namely the market
may be efficient when the peaks appear.
 
The author thanks Prof. Dr. T. Munakata for stimulative discussions and 
useful comments. This work is partially supported by the Japan Society 
for the Promotion of Science, Grant-in-Aid for Scientific Research \#
17760067.

\end{document}